\documentclass[12pt]{article}
\usepackage{graphicx}
\begin{document}

\vbox{\vspace{5ex}}

\begin{center}
{\Large \bf The Question of Simultaneity in Relativity and Quantum Mechanics}

\vspace{3ex}

Y. S. Kim \footnote{electronic address: yskim@physics.umd.edu}\\
Department of Physics, University of Maryland,\\
College Park, Maryland 20742, U.S.A.\\

\vspace{3ex}

Marilyn E. Noz \footnote{electronic address: noz@nucmed.med.nyu.edu}\\
Department of Radiology, New York University,\\ New York, New York 10016, U.S.A.\\

\end{center}

\vspace{3ex}

\begin{abstract}
In relativity, two simultaneous events at two different places
are not simultaneous for observers in different Lorentz frames.  In
the Einstein-Podolsky-Rosen experiment, two simultaneous measurements
are taken at two different places.  Would they still be simultaneous
to observers in moving frames?  It is a difficult question, but it is
still possible to study this problem in the microscopic world. In the
hydrogen atom, the incertainty can be considered to be entirely
associated with the ground-state.  However, is there an uncertainty
associated with the time-separation variable between the proton and
electron?  This time-separation variable is a forgotten, if not hidden,
variable in the present form of quantum mechanics.  The first step toward
the simultaneity problem is to study the role of this time-separation
variable in the Lorentz-covariant world.  It is shown possible to study
this problem using harmonic oscillators applicable to hadrons which are
bound states of quarks.  It is also possible to derive consequences
that can be tested experimentally.
\end{abstract}

\section{Introduction}\label{intro}

In his book entitled "Encounters with Einstein"~\cite{heisen83},
Heisenberg states that the mathematics of Lorentz transformations was
easy to understand and appreciate, but the concept of simultaneity in
Einstein's relativity was difficult to grasp.
Heisenberg had this problem before he formulated his uncertainty
relation, and the concept of simultaneity plays a pivotal role in the
interpretation there.  Is Heisenberg's simultaneity consistent with
Einstein's simultaneity?

When we talk about simultaneous measurements, we are uncritical
about whether those measurements are taken at the same place or
different places.  In the EPR-type experiments~\cite{epr35},
two simultaneous measurements are taken at two different places.
Would these two measurements appear simultaneous to an observer
on a bicycle?  We do not know where the story stands on this issue,
because the problem includes both macroscopic and microscopic
scales.  This involves localization problems, in addition to the
simultaneity issue.  We are not able to provide a resolution to
this problem in the present report.

On the other hand, we can study the problem in the microscopic scale.
The radius of the ground-state hydrogen atom can be regarded as an
uncertain quantity.  There is a spacial separation between between
the proton and electron.  These two particles are located at
different places.  We have to measure the position of the proton
and that of the electron to take the difference.  In the present
form of quantum mechanics, we assume that they are taken
simultaneously because we never worry about the time separation
between them.  If we believe in Einstein's relativity, there is
necessarily a time-separation variable between these two particles,
and it will play a prominent role for observers in different
Lorentz frames.

In order to approach the problem, let us go back to Heisenberg's
problem.  It is easier to understand the mathematics of the Lorentz
transformation than the concept of simultaneity.  It would thus
be easier if we build the mathematical framework first.  We may then
be able to give physical interpretations, and also derive consequences
derivable from the mathematical formalism.  The modern version of
the hydrogen atom is a bound-state of quarks, called the hadron.  While
there are no experimental data on hydrogen atoms moving with relativistic
speed, the physics of hadrons involves bound states in the
Lorentz-covariant world.

According to our experience, the present form of quantum mechanics
is largely a physics of harmonic oscillators.  Since the group
consisting of two-by-two unimodular matrices, or $SL(2,C)$, forms the
universal covering group of the Lorentz group, special relativity is a
physics of two-by-two matrices.  Therefore, the coupled harmonic
oscillator can provide a concrete model for relativistic quantum
mechanics.

With this point in mind, Dirac and Feynman used harmonic oscillators
to test their physical ideas.  In this paper, we first examine Dirac's
attempts to combine quantum mechanics with relativity in his own style:
to construct mathematically appealing models.  We then examine how
Feynman approached this problem.  He insisted on his own style:
observe the experimental world, tell the story of the real world, and
then write down mathematical formulas as needed.

In this paper, we use coupled harmonic oscillators to build a bridge
between the two different attempts made by Dirac and Feynman.  The
coupled oscillator system not only connects the ideas of these two
great physicists, but also serves as an illustrative tool for some
of the current ideas in physics, such as entanglement and decoherence.

As for observable consequences of the oscillator formalism which connects
Dirac and Feynman, we would like to discuss in this report Feynman's
parton picture which is valid in the Lorentz frame in which hadronic
speed is close to that of light.  It is widely believed that hadrons
are bound states of quarks like the hydrogen atom when they are at rest.
Then why are partons so different from the quarks inside the static
hadron?  We shall discuss how the time-separation variable plays the
crucial role in resolving this quark-parton puzzle.

In Sec.~\ref{quantu}, we start with the classical Hamiltonian for two
coupled oscillators.  It is possible to obtain a explicit solution for
the Schr\"odinger equation in terms of the normal coordinates.  We
then derive a convenient form of this solution from which the concept
of entanglement can be studied thoroughly.
Section~\ref{dirosc} examines Dirac's life-long attempt to combine
quantum mechanics with special relativity.  In Sec.~\ref{adden}, we
study some of the problems which Dirac left us to solve.
In Sec.~\ref{feyosc},
We construct a covariant model of
relativistic extended particles by combining Dirac's oscillators
with Feynman's phenomenological approach to relativistic quark model.
It is shown that Feynman's parton model can be interpreted as a limiting
case of one covariant model for a covariant bound-state model.

\section{Coupled Oscillators and Entangled Oscillators}\label{quantu}

Two coupled harmonic oscillators serve many different purposes in
physics.  It is well known that this oscillator problem can be
formulated into a problem of a quadratic equation in two variables.
The diagonalization of the quadratic form includes a rotation of the
coordinate system.  However, the diagonalization process requires
additional transformations involving the scales of the coordinate
variables~\cite{arav89,hkn99ajp}.  Indeed, it was found that the
mathematics of this procedure can be as complicated as the group
theory of Lorentz transformations in a six dimensional space with
three spatial and three time coordinates~\cite{hkn95jm}.

However, in this paper, we start with a simple problem of two oscillators
with equal mass.  This contains enough physics for our present purpose.
Then the Hamiltonian takes the form
\begin{equation}\label{eq.1}
H = {1\over 2}\left\{{1\over m} p^{2}_{1} + {1\over m}p^{2}_{2}
+ A x^{2}_{1} + A x^{2}_{2} + 2C x_{1} x_{2} \right\}.
\end{equation}

If we choose coordinate variables
\begin{eqnarray} \label{eq.3}
&{}& y_{1} = {1\over\sqrt{2}}\left(x_{1} + x_{2}\right) , \nonumber\\[2ex]
&{}& y_{2} = {1\over\sqrt{2}}\left(x_{1} - x_{2}\right) ,
\end{eqnarray}
the Hamiltonian can be written as
\begin{equation}\label{eq.6}
H = {1\over 2m} \left\{p^{2}_{1} + p^{2}_{2} \right\} +
{K\over 2}\left\{e^{-2\eta} y^{2}_{1} + e^{2\eta} y^{2}_{2} \right\} ,
\end{equation}
where
\begin{eqnarray}\label{eq.5}
&{}&   K = \sqrt{A^{2} - C^{2}} ,  \nonumber \\[.5ex]
&{}& \exp(2\eta) =\sqrt{\frac{A - C}{A + C} } ,
\end{eqnarray}
The classical eigenfrequencies are $\omega_{\pm} = \omega e^{\pm}$ with
\begin{equation}\label{omega}
\omega = \sqrt{\frac{K}{m}} .
\end{equation}

If $y_{1}$ and $y_{2}$ are measured in units of $(mK)^{1/4} $,
the ground-state wave function of this oscillator system is
\begin{equation}\label{eq.13}
\psi_{\eta}(x_{1},x_{2}) = {1 \over \sqrt{\pi}}
\exp{\left\{-{1\over 2}(e^{-\eta} y^{2}_{1} + e^{\eta} y^{2}_{2})
\right\} } ,
\end{equation}
The wave function is separable in the $y_{1}$ and $y_{2}$ variables.
However, for the variables $x_{1}$ and $x_{2}$, the story is quite
different, and can be extended to the issue of entanglement.

There are three ways to excite this ground-state oscillator system.
One way is to multiply Hermite polynomials for the usual quantum
excitations.  The second way is to construct coherent states for
each of the $y$ variables.  Yet, another way is to construct
thermal excitations.  This requires density matrices and Wigner
functions~\cite{hkn99ajp}.

The key question is how the quantum mechanics in the world of the
$x_{1}$ variable is affected by the $x_{2}$ variable.  If the
$x_{2}$ space is not observed, it corresponds to Feynman's rest
of the universe.  If we use two separate measurement processes for
these two variables, these two oscillators are  entangled.

Let us write the wave function of Eq.(\ref{eq.13}) in terms of
$x_{1}$ and $x_{2}$, then
\begin{equation}\label{eq.14}
\psi_{\eta}(x_{1},x_{2}) = {1 \over \sqrt{\pi}}
\exp\left\{-{1\over 4}\left[e^{-\eta}(x_{1} + x_{2})^{2} +
e^{\eta}(x_{1} - x_{2})^{2} \right] \right\} .
\end{equation}
When the system is decoupled with $\eta = 0$, this wave function becomes
\begin{equation}\label{eq.15}
\psi_{0}(x_{1},x_{2}) = \frac{1}{\sqrt{\pi}}
\exp{\left\{-{1\over 2}(x^{2}_{1} + x^{2}_{2}) \right\}} .
\end{equation}
The system becomes separable and becomes disentangled.

As was discussed in the literature for several different
purposes~\cite{kno79ajp,knp86,knp91}, this wave function can be
expanded as
\begin{equation}\label{expan}
\psi_{\eta }(x_{1},x_{2}) = {1 \over \cosh\eta}\sum^{}_{k}
\left(\tanh{\eta \over 2}\right)^{k} \phi_{k}(x_{1}) \phi_{k}(x_{2}) ,
\end{equation}
where $\phi_{k}(x)$ is the harmonic oscillator wave function for the
$k-th$ excited state.
This expansion serves as the mathematical basis for squeezed states
of light in quantum optics~\cite{knp91}, among other applications.

In addition, this expression clearly demonstrates that coupled
oscillators are entangled oscillators.  Let us look at the expression
of Eq.(\ref{expan}).  If the variable $x_{1}$ and $x_{2}$ are measured
separately.

In Sec~\ref{dirosc}, we shall see that the mathematics of coupled
oscillators can serve as the basis for the covariant harmonic
oscillator formalism where the $x_{1}$ and $x_{2}$ variables
are replaced by the longitudinal and time-like variables,
respectively.  This mathematical identity will leads to the
concept of space-time entanglement in special relativity.

\section{Dirac's Harmonic Oscillators}\label{dirosc}

Paul A. M. Dirac is known to us through the Dirac equation for spin-1/2
particles.  But his main interest was in foundational problems.
First, Dirac was never satisfied with the probabilistic formulation of
quantum mechanics.  This is still one of the hotly debated subjects in
physics.  Second, if we tentatively accept the present form
of quantum mechanics, Dirac insisted that it had to be consistent
with special relativity.  He wrote several important papers on this
subject.  Let us look at some of his papers on this subject.

\begin{figure}[thb]
\centerline{\includegraphics[scale=0.8]{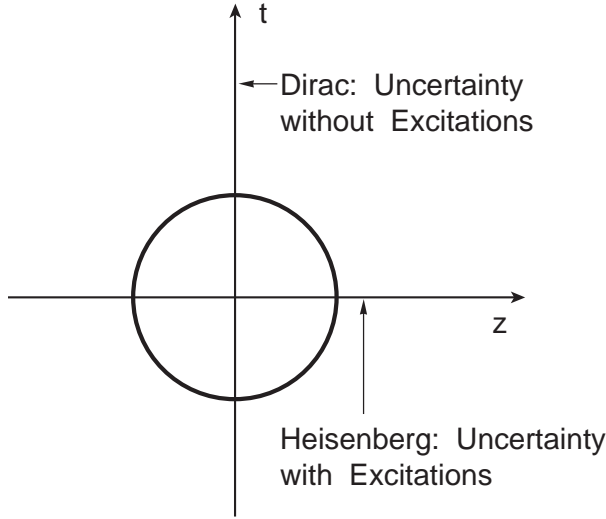}}
\vspace{5mm}
\caption{Space-time picture of quantum mechanics.  There
are quantum excitations along the space-like longitudinal direction, but
there are no excitations along the time-like direction.  The time-energy
relation is a c-number uncertainty relation.}\label{quantum}
\end{figure}

During World War II, Dirac was looking into the possibility of constructing
representations of the Lorentz group using harmonic oscillator wave
functions~\cite{dir45}.  The Lorentz group is the language of special
relativity, and the present form of quantum mechanics starts with harmonic
oscillators.  Presumably, therefore, he was interested in making quantum
mechanics Lorentz-covariant by constructing representations of the Lorentz
group using harmonic oscillators.

In his 1945 paper~\cite{dir45}, Dirac considers the Gaussian form
\begin{equation}
\exp\left\{- {1 \over 2}\left(x^2 + y^2 + z^2 + t^2\right)\right\} .
\end{equation}
We note that this Gaussian form is in the $(x,~y,~z,~t)$
coordinate variables.  Thus, if we consider a Lorentz boost along the
$z$ direction, we can drop the $x$ and $y$ variables, and write the
above equation as
\begin{equation}\label{ground}
\exp\left\{- {1 \over 2}\left(z^2 + t^2\right)\right\} .
\end{equation}
This is a strange expression for those who believe in Lorentz invariance.
The expression
\begin{equation}
\exp\left\{- {1 \over 2}\left(z^2 - t^2\right)\right\} .
\end{equation}
is invariant, but Dirac's Gaussian form of Eq.(\ref{ground}) is not.

On the other hand, this expression is consistent with his earlier papers
on the time-energy uncertainty relation~\cite{dir27}.  In those papers,
Dirac observes that there is a time-energy uncertainty relation, while
there are no excitations along the time axis.  He called this the
``c-number time-energy uncertainty'' relation.  When one of us
(YSK) was talking with Dirac in 1978, he clearly mentioned
this word again.  He said further that this is one of the stumbling
block in combining quantum mechanics with relativity.  This
situation is illustrated in Fig.~\ref{quantum}.

\begin{figure}[thb]
\centerline{\includegraphics[scale=0.8]{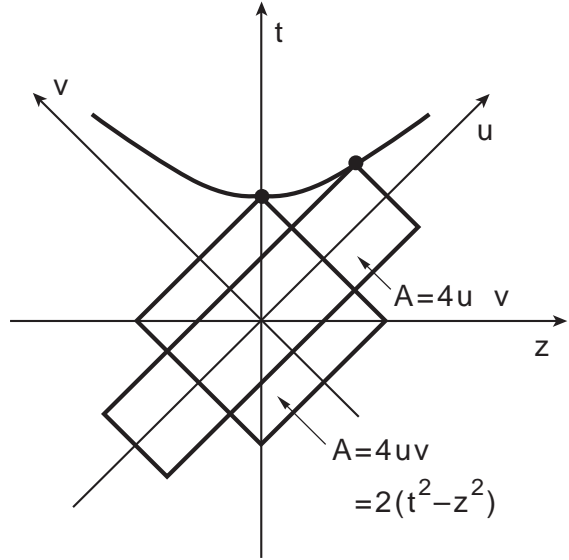}}
\vspace{5mm}
\caption{Lorentz boost in the light-cone coordinate
system.}\label{licone}
\end{figure}

Let us look at Fig.~\ref{quantum} carefully.  This figure is a pictorial
representation of Dirac's Eq.(\ref{ground}),  with localization in both
space and time coordinates.  Then Dirac's fundamental question would be
how to make this figure covariant?  This is where Dirac stops.  However,
this is not the end of the Dirac story.
\begin{figure}[thb]
\centerline{\includegraphics[scale=0.4]{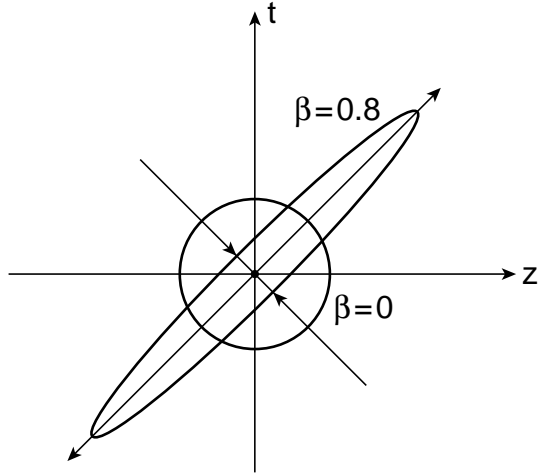}}
\caption{Effect of the Lorentz boost on the space-time
wave function.  The circular space-time distribution in the rest frame
becomes Lorentz-squeezed to become an elliptic
distribution.}\label{ellipse}
\end{figure}

Dirac's interest in harmonic oscillators did not stop with his 1945
paper on the representations of the Lorentz group.  In his
1963~\cite{dir63} paper, he constructed a representation
of the $O(3,2)$ deSitter group using two coupled harmonic oscillators.
This paper contains not only the mathematics of combining special
relativity with the quantum mechanics of quarks inside hadrons, but
also forms the foundations of two-mode squeezed states which are so
essential to modern quantum optics~\cite{knp91}.   Dirac did not know
this when he was writing his 1963 paper.

Furthermore, the $O(3,2)$ deSitter group contains the Lorentz group
$O(3,1)$ as a subgroup.  Thus, Dirac's oscillator representation of
the deSitter group essentially contains all the mathematical
ingredient of what we are doing in this paper.

\section{Addendum to Dirac'c Oscillators}\label{adden}

In 1949, the Reviews of Modern Physics published a special issue to
celebrate Einstein's 70th birthday.  This issue contains Dirac paper
entitled ``Forms of Relativistic Dynamics''~\cite{dir49}.
In this paper, he introduced his light-cone coordinate system,
in which a Lorentz boost becomes a squeeze transformation.

When the system is boosted along the $z$ direction, the transformation
takes the form
\begin{equation}\label{boostm}
\pmatrix{z' \cr t'} = \pmatrix{\cosh(\eta/2) & \sinh(\eta/2) \cr
\sinh(\eta/2) & \cosh(\eta/2) } \pmatrix{z \cr t} .
\end{equation}

This is not a rotation, and people still feel strange about this
form of transformation.  In 1949~\cite{dir49}, Dirac introduced his
light-cone
variables defined as~\cite{dir49}
\begin{equation}\label{lcvari}
u = (z + t)/\sqrt{2} , \qquad v = (z - t)/\sqrt{2} ,
\end{equation}
the boost transformation of Eq.(\ref{boostm}) takes the form
\begin{equation}\label{lorensq}
u' = e^{\eta/2 } u , \qquad v' = e^{-\eta/2 } v .
\end{equation}
The $u$ variable becomes expanded while the $v$ variable becomes
contracted, as is illustrated in Fig.~\ref{licone}.  Their product
\begin{equation}
uv = {1 \over 2}(z + t)(z - t) = {1 \over 2}\left(z^2 - t^2\right)
\end{equation}
remains invariant.  In Dirac's picture, the Lorentz boost is a
squeeze transformation.

If we combine Fig.~\ref{quantum} and Fig.~\ref{licone}, then we end up
with Fig.~\ref{ellipse}.
In mathematical formulae, this transformation changes the Gaussian form
of Eq.(\ref{ground}) into
\begin{equation}\label{eta}
\psi_{\eta }(z,t) = \left({1 \over \pi }\right)^{1/2}
\exp\left\{-{1\over 2}\left(e^{-\eta }u^{2} +
e^{\eta}v^{2}\right)\right\} .
\end{equation}
Let us go back to Sec.~\ref{quantu} on the coupled oscillators.  The
above expression is the same as Eq.(\ref{eq.13}).  The $x_{1}$ variable
now became the longitudinal variable $z$, and the $x_{2}$ variable
became the time like variable $t$.

We can use coupled harmonic oscillators as the starting point of
relativistic quantum mechanics.  This allows us to translate the quantum
mechanics of two coupled oscillators defined over the space of
$x_{1}$ and $x_{2}$ into the quantum mechanics defined over the
space time region of $z$ and $t$.

This form becomes (\ref{ground}) when $\eta$ becomes zero.  The
transition from Eq.(\ref{ground}) to Eq.(\ref{eta}) is a squeeze
transformation.
It is now possible to combine what Dirac observed into a covariant formulation
of the harmonic oscillator system. First, we can combine his c-number
time-energy uncertainty relation described in Fig.~\ref{quantum}
and his light-cone coordinate system of Fig.~\ref{licone} into
a picture of covariant space-time localization given in
Fig.~\ref{ellipse}.

In addition, there are two more homework problems which Dirac left
us to solve. First, in defining the $t$ variable for the Gaussian form of
Eq.(\ref{ground}),  Dirac did not specify the physics of this variable.
If it is going to be the calendar time, this form vanishes in the remote
past and remote future.  We are not dealing with this kind of object in
physics.  What is then the physics of this time-like $t$ variable?

\begin{figure}[thb]
\centerline{\includegraphics[scale=0.7]{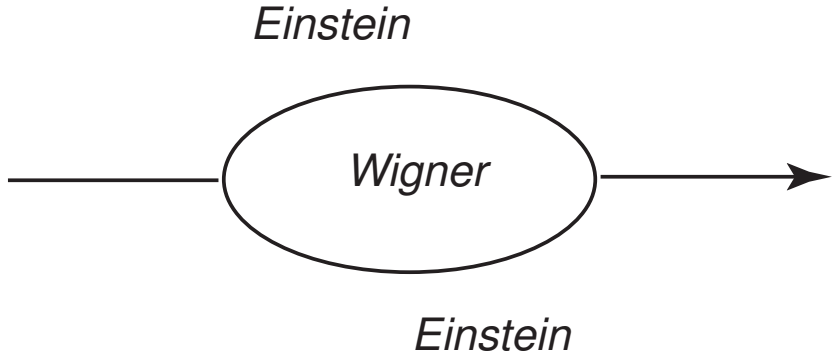}}
\vspace{5mm}
\caption{Wigner in Einstein's world.  Einstein formulates special
relativity whose energy-momentum relation is valid for point particles
as well as particles with internal space-time structure.  It was Wigner
who formulated the framework for internal space-time symmetries by introducing his
little groups whose transformations leave the four-momentum of a given particle
invariant.}\label{dff22}
\end{figure}
The Schr\"odinger quantum mechanics of the hydrogen atom deals with
localized probability distributions.  Indeed, the localization condition
leads to the discrete energy spectrum.  Here, the uncertainty relation
is stated in terms of the spatial separation between the proton and
the electron.  If we believe in Lorentz covariance, there must also
be a time-separation between the two constituent particles, and an
uncertainty relation applicable to this separation variable.  Dirac
did not say in his papers of 1927 and 1945, but Dirac's ``t'' variable
is applicable to this time-separation variable.  This time-separation
variable will be discussed in detail in Sec.~\ref{feyosc} for the
case of relativistic extended particles.

Second, as for the time-energy uncertainty relation, Dirac'c concern
was how the c-number time-energy uncertainty relation without excitations
can be combined with uncertainties in the position space with excitations.
Dirac's 1927 paper was written before Wigner's 1939 paper on the internal
space-time symmetries of relativistic particles.

Both of these questions can be answered in terms of the space-time
symmetry of bound states in the Lorentz-covariant regime.  In his
1939 paper, Wigner worked out internal space-time symmetries of
relativistic particles.  He approached the problem by constructing
the maximal subgroup of the Lorentz group whose transformations leave
the given four-momentum invariant.  As a consequence, the internal
symmetry of a massive particle is like the three-dimensional rotation
group.  This is shown in Fig.~\ref{dff22}.

If we extend this concept to relativistic bound states, the space-time
asymmetry which Dirac observed in 1927 is quite consistent with Einstein's
Lorentz covariance.  The time variable can be treated separately.
Furthermore, it is possible to construct a representations of Wigner's
little group for massive particles~\cite{knp86}.
As for the time-separation, it is also a variable governing
internal space-time symmetry which can be linearly mixed when the
system is Lorentz-boosted.

\section{Feynman's Oscillators }\label{feyosc}

\begin{figure}[thb]
\centerline{\includegraphics[scale=0.7]{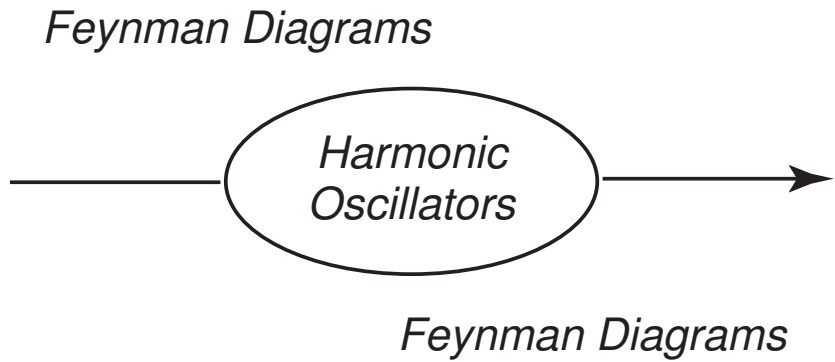}}
\vspace{5mm}
\caption{Feynman's roadmap for combining quantum mechanics with special
relativity.  Feynman diagrams work for running waves, and they provide
a satisfactory resolution for scattering states in Einstein's world.
For standing waves trapped inside an extended hadron, Feynman suggested
harmonic oscillators as the first step.}\label{dff33}
\end{figure}

Quantum field theory has been quite successful in terms of Feynman
diagrams based on the S-matrix formalism, but is useful only for physical
processes where a set of free particles becomes another set of free
particles after interaction.  Quantum field theory does not address the
question of localized probability distributions and their covariance
under Lorentz transformations.  In order to address this question,
Feynman {\it et al.} suggested harmonic oscillators to tackle the
problem~\cite{fkr71}.  Their idea is indicated in Fig.~\ref{dff33}.

Before 1964~\cite{gell64}, the hydrogen atom was used for
illustrating bound states.  These days, we use hadrons which are
bound states of quarks.  Let us use the simplest hadron consisting of
two quarks bound together with an attractive force, and consider their
space-time positions $x_{a}$ and $x_{b}$, and use the variables
\begin{equation}
X = (x_{a} + x_{b})/2 , \qquad x = (x_{a} - x_{b})/2\sqrt{2} .
\end{equation}
The four-vector $X$ specifies where the hadron is located in space and
time, while the variable $x$ measures the space-time separation
between the quarks.  According to Einstein, this space-time separation
contains a time-like component which actively participates as in
Eq.(\ref{boostm}), if the hadron is boosted along the $z$ direction.
This boost can be conveniently described by the light-cone variables
defined in Eq(\ref{lcvari}).
Does this time-separation variable exist when the hadron is at rest?
Yes, according to Einstein.  In the present form of quantum mechanics,
we pretend not to know anything about this variable.  Indeed, this
variable belongs to Feynman's rest of the universe.

What do Feynman {\it et al.} say about this oscillator wave function?
In their classic 1971 paper~\cite{fkr71}, Feynman {\it et al.} start
with the following Lorentz-invariant differential equation.
\begin{equation}\label{osceq}
{1\over 2} \left\{x^{2}_{\mu} -
{\partial^{2} \over \partial x_{\mu }^{2}}
\right\} \psi(x) = \lambda \psi(x) .
\end{equation}
This partial differential equation has many different solutions
depending on the choice of separable variables and boundary conditions.
Feynman {\it et al.} insist on Lorentz-invariant solutions which are
not normalizable.  On the other hand, if we insist on normalization,
the ground-state wave function takes the form of Eq.(\ref{ground}).
It is then possible to construct a representation of the
Poincar\'e group from the solutions of the above differential
equation~\cite{knp86}.  If the system is boosted, the wave function
becomes given in Eq.(\ref{eta}).

\begin{figure}
\centerline{\includegraphics[scale=0.5]{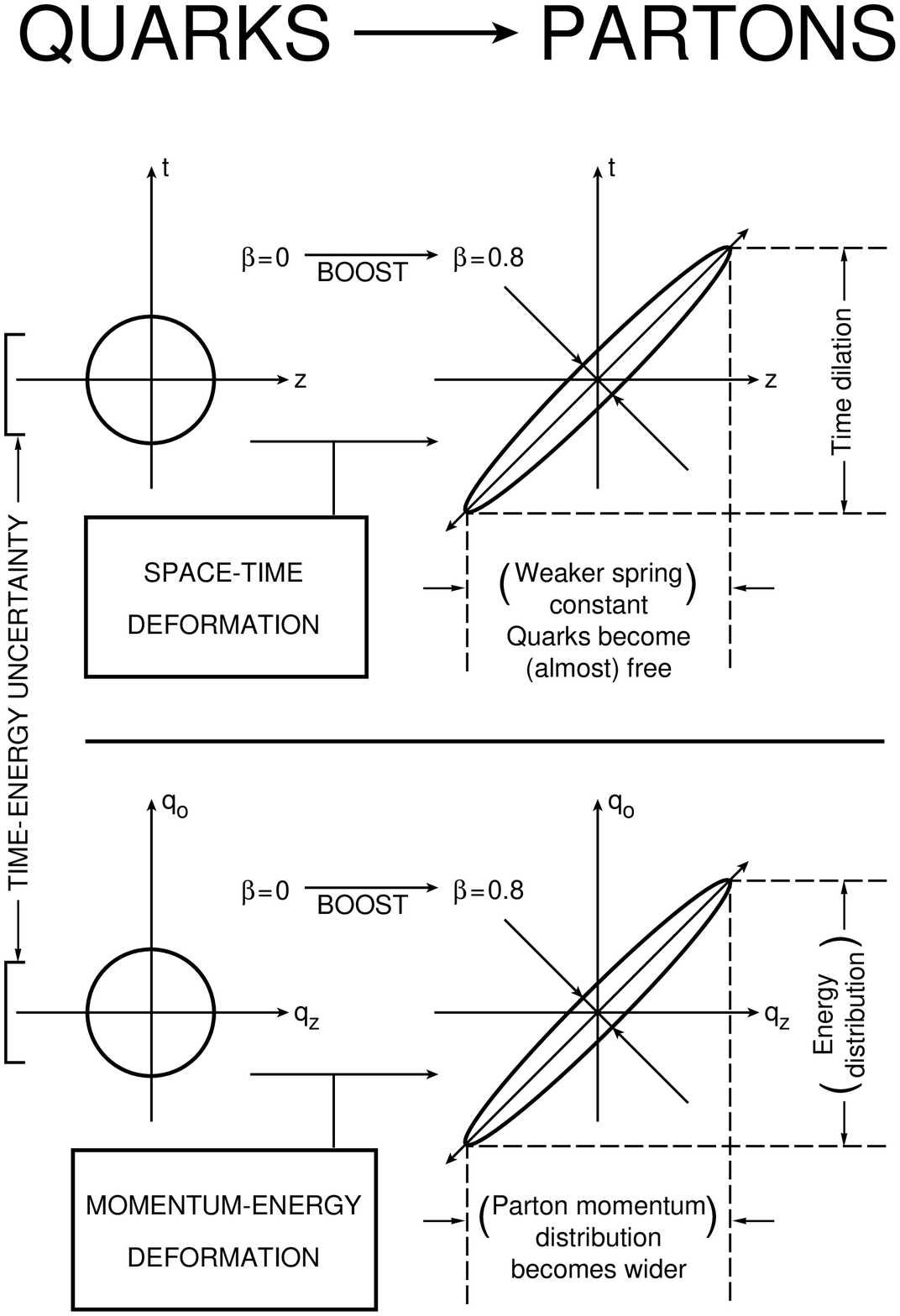}}
\vspace{5mm}
\caption{Lorentz-squeezed space-time and momentum-energy wave
functions.  As the hadron's speed approaches that of light, both
wave functions become concentrated along their respective positive
light-cone axes.  These light-cone concentrations lead to Feynman's
parton picture.}\label{parton}
\end{figure}

This wave function becomes Eq.(\ref{ground}) if $\eta$ becomes zero.
The transition from Eq.(\ref{ground}) to Eq.(\ref{eta}) is a
squeeze transformation.  The wave function of Eq.(\ref{ground}) is
distributed within a circular region in the $u v$ plane, and thus
in the $z t$ plane.  On the other hand, the wave function of
Eq.(\ref{eta}) is distributed in an elliptic region with the light-cone
axes as the major and minor axes respectively.  If $\eta$ becomes very
large, the wave function becomes concentrated along one of the
light-cone axes.  Indeed, the form given in Eq.(\ref{eta}) is a
Lorentz-squeezed wave  function.  This squeeze mechanism is
illustrated in Fig.~\ref{ellipse}.

There are many different solutions of the Lorentz invariant differential
equation of Eq.(\ref{osceq}).  The solution given in Eq.(\ref{eta})
is not Lorentz invariant but is covariant.  It is normalizable in the
$t$ variable, as well as in the space-separation variable $z$.  It is
indeed possible to construct Wigner's $O(3)$-like little group for massive
particles~\cite{wig39}, and thus the representation of the Poincar\'e
group~\cite{knp86}.  Our next question is whether this formalism has
anything to do with the real world.

\begin{figure}
\centerline{\includegraphics[scale=0.6]{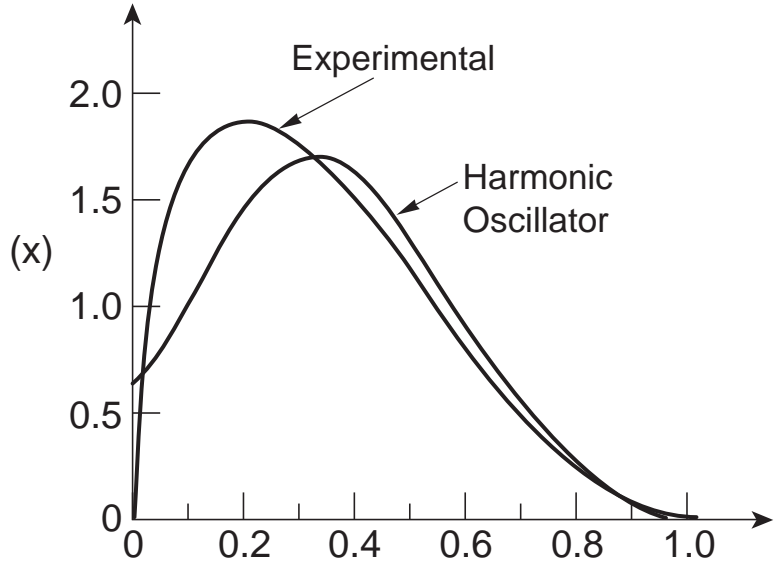}}
\vspace{5mm}
\caption{Parton distribution function.
Theory and experiment.}\label{hussar}
\end{figure}

In 1969, Feynman observed that a fast-moving hadron can be regarded
as a collection of many ``partons'' whose properties appear to be
quite different from those of the quarks~\cite{fey69}.  For example,
the number of quarks inside a static proton is three, while the number
of partons in a rapidly moving proton appears to be infinite.  The
question then is how the proton looking like a bound state of quarks
to one observer can appear different to an observer in a different
Lorentz frame?  Feynman made the following systematic observations.

\begin{itemize}

\item[a.]  The picture is valid only for hadrons moving with
  velocity close to that of light.

\item[b.]  The interaction time between the quarks becomes dilated,
   and partons behave as free independent particles.

\item[c.]  The momentum distribution of partons becomes widespread as
   the hadron moves fast.

\item[d.]  The number of partons seems to be infinite or much larger
    than that of quarks.

\end{itemize}

\noindent Because the hadron is believed to be a bound state of two
or three quarks, each of the above phenomena appears as a paradox,
particularly b) and c) together.

In order to resolve this paradox, let us write down the
momentum-energy wave function corresponding to Eq.(\ref{eta}).
If we let the quarks have the four-momenta $p_{a}$ and $p_{b}$, it is
possible to construct two independent four-momentum
variables~\cite{fkr71}
\begin{equation}
P = p_{a} + p_{b} , \qquad q = \sqrt{2}(p_{a} - p_{b}) ,
\end{equation}
where $P$ is the total four-momentum.  It is thus the hadronic
four-momentum.

The variable $q$ measures the four-momentum separation between
the quarks.  Their light-cone variables are
\begin{equation}\label{conju}
q_{u} = (q_{0} - q_{z})/\sqrt{2} ,  \qquad
q_{v} = (q_{0} + q_{z})/\sqrt{2} .
\end{equation}
The resulting momentum-energy wave function is
\begin{equation}\label{phi}
\phi_{\eta }(q_{z},q_{0}) = \left({1 \over \pi }\right)^{1/2}
\exp\left\{-{1\over 2}\left(e^{\eta}q_{u}^{2} +
e^{-\eta}q_{v}^{2}\right)\right\} .
\end{equation}
Because we are using here the harmonic oscillator, the mathematical
form of the above momentum-energy wave function is identical to that
of the space-time wave function.  The Lorentz squeeze properties of
these wave functions are also the same.  This aspect of the squeeze
has been exhaustively discussed in the
literature~\cite{knp86,kn77par,kim89}.

When the hadron is at rest with $\eta = 0$, both wave functions
behave like those for the static bound state of quarks.  As $\eta$
increases, the wave functions become continuously squeezed until
they become concentrated along their respective positive
light-cone axes.  Let us look at the z-axis projection of the
space-time wave function.  Indeed, the width of the quark distribution
increases as the hadronic speed approaches that of the speed of
light.  The position of each quark appears widespread to the observer
in the laboratory frame, and the quarks appear like free particles.

The momentum-energy wave function is just like the space-time wave
function, as is shown in Fig.~\ref{parton}.  The longitudinal momentum
distribution becomes wide-spread as the hadronic speed approaches the
velocity of light.  This is in contradiction with our expectation from
non-relativistic quantum mechanics that the width of the momentum
distribution is inversely proportional to that of the position wave
function.  Our expectation is that if the quarks are free, they must
have sharply defined momenta, not a wide-spread distribution.

However, according to our Lorentz-squeezed space-time and
momentum-energy wave functions, the space-time width and the
momentum-energy width increase in the same direction as the hadron
is boosted.  This is of course an effect of Lorentz covariance.
This indeed is the key to the resolution of the quark-parton
paradox~\cite{knp86,kn77par}.

After these qualitative arguments, we are interested in whether
Lorentz-boosted bound-state wave functions in the hadronic rest
frame could lead to parton distribution functions.  If we start with
the ground-state Gaussian wave function for the three-quark wave
function for the proton, the parton distribution function appears
as Gaussian as is indicated in Fig.~\ref{hussar}.  This Gaussian  form
is compared with experimental distribution also in Fig.~\ref{hussar}.

For large $x$ region, the agreement is excellent, but the agreement is
not satisfactory for small values of $x$.  In this region, there is
a complication called the ``sea quarks.''  However, good sea-quark physics
starts from good valence-quark physics.  Figure~\ref{hussar} indicates
that the boosted ground-state wave function provides a good valence-quark
physics.

Feynman's parton picture is one of the most controversial models
proposed in the 20th century.  The original model is valid only in
Lorentz frames where the initial proton moves with infinite momentum.
It is gratifying to note that this model can be produced as a limiting
case of one covariant model which produces the quark model in the
frame where the proton is at rest.

\section*{Concluding Remarks}\label{concl}
In this report, we considered Einstein's relative simultaneity working
in the wave-function picture of quantum mechanics.   It was shown that
the covariant harmonic oscillator applicable to hadrons in the quark
model gives an illustration of how this problem can be approached.  A
non-zero spacial-separation in the hadronic rest frame with zero
time-separation gives a measurable effect in the Lorentz frame
which moves with a velocity close that of light.

\begin{figure}[thb]
\centerline{\includegraphics[scale=0.5]{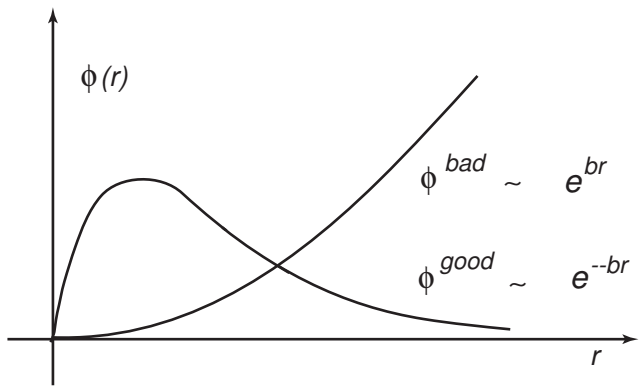}}
\vspace{5mm}
\caption{Good and bad wave functions contained in the S-matrix.
Bound-state wave functions satisfy the localization condition and are
good wave functions.  Analytic continuations of plane waves do not
satisfy the localization boundary condition, and become bad wave
functions at the bound-state
energy.}\label{goodbad}
\end{figure}

It is widely understood that the present form of quantum field theory,
with the S-matrix and Feynman diagrams, does the job of combining quantum
mechanics and relativity.  The question then is why we did not use field
theory to deal with the simultaneity problem.  The answer is very simple.
The present
form of field theory can deal only with scattering problems.  There have
been many attempts in the past to extend the field theory algorithm to
bound state problems.  This requires analytic continuation of incoming
and outgoing waves to negative energy regions.  The outgoing waves
become localized bound-state wave functions, but the incoming waves
increase exponentially for large distances.  We do not know how to
control this localization problem in quantum field theory, as we can see
from the Dashen-Frautschi fiasco~\cite{df64,kim66}.  Feynman
was right.  We should start bound-state problems with localized
harmonic oscillator wave functions.

Indeed, the once-celebrated calculation of the neutron-proton mass
difference by Dashen and Frautschi illustrates difficulties of using
the present form of field theory for bound state problems~\cite{df64}.
In order to calculate the mass difference as an electromagnetic
perturbation, they developed a perturbation formula solely based on
the S-matrix quantities, but they ended up with a first-order energy shift
corresponding to~\cite{kim66}
\begin{equation}\label{shift}
\delta E = \left(\phi^{good}, \delta V \phi^{bad} \right) ,
\end{equation}
where the good and bad bound-state wave functions are like
\begin{equation}\label{wfs}
 \phi^{good} \sim e^{-br} , \qquad
\phi^{bad} \sim e^{br} ,
\end{equation}
for large values of $r$, as illustrated in Fig.~\ref{goodbad}.  We are
not aware of any S-matrix or field theoretic method which guarantees
the localization of bound-state wave functions.

\end{document}